\documentclass[letterpaper,twocolumn,10pt]{article}
\PassOptionsToPackage{hyphens}{url} 
\usepackage{template_usenix/usenix-2020-09}

\usepackage{tikz}
\usetikzlibrary{tikzmark}
\usetikzlibrary{decorations.pathreplacing}
\usetikzlibrary{backgrounds}

\usepackage{forest}
\usepackage{dirtree}
\usepackage{hyperref}
\usepackage{soul}
\usepackage{comment}
\usepackage{booktabs} 
\usepackage{wasysym} 
\usepackage{multirow} 
\usepackage{float} 
\usepackage{arydshln} 
\usepackage{dblfloatfix} 

\usepackage{titlesec}

\usepackage[inline]{enumitem}
\setlist{itemsep=.5ex}

\microtypecontext{spacing=nonfrench}

\newcommand\todo[1]{\textcolor{red}{\ul{#1}}}
\newcommand\trent[1]{\textcolor{blue}{\ul{#1}}}

\newcommand\zhiyun[1]{\textcolor{orange}{\ul{Zhiyun: #1}}}

\newcommand\git{{\tt git}}
\newcommand\httpd{{\tt httpd}}
\newcommand\tar{{\tt tar}}

\newcommand{\f}[1]{{\tt #1}} 

\begin{document}

\title{Unsafe at Any Copy: Name Collisions from Mixing Case Sensitivities}

\author{~}

\newcommand\psu{\textsuperscript{$*$}}
\newcommand\ucriverside{\textsuperscript{\dag}}

\author{
   {\rm Aditya Basu\psu}\\
   aditya.basu@psu.edu
   	\and
   {\rm John Sampson\psu}\\
   jms1257@psu.edu
   	\and
   {\rm Zhiyun Qian\ucriverside}\\
   zhiyunq@cs.ucr.edu
   	\and
   {\rm Trent Jaeger\psu}\\
   trj1@psu.edu
   	\and
   {\psu The Pennsylvania State University}\\
   {\ucriverside University of California, Riverside}
} 


\maketitle

\begin{abstract}
File name confusion attacks, such as malicious symlinks and file squatting, have long been studied as sources of security vulnerabilities. However, a recently emerged type, i.e., \textit{\textbf{case-sensitivity-induced name collisions}}, has not been scrutinized.
These collisions are introduced by differences in name resolution under case-sensitive and case-insensitive file systems or directories.
A prominent example is the recent Git vulnerability (CVE-2021-21300) which can lead to code execution on a victim client when it clones a maliciously crafted repository onto a case-insensitive file system.
With trends including {\tt ext4} adding support for per-directory case-insensitivity and the broad deployment of the Windows Subsystem for Linux, the prerequisites for such vulnerabilities are increasingly likely to exist even in a single system.

In this paper, we make a first effort to investigate how and where the lack of any uniform approach to handling name collisions leads to a diffusion of responsibility and resultant vulnerabilities. 
Interestingly, we demonstrate the existence of a range of novel security challenges arising from name collisions and their inconsistent handling by low-level utilities and applications. Specifically, our experiments show that utilities handle many name collision scenarios unsafely, leaving the responsibility to applications whose developers are unfortunately not yet aware of the threats. 
We examine three case studies as a first step towards systematically understanding the emerging type of name collision vulnerability. 
\end{abstract}

\section{Introduction}
\label{sec:intro}


A fundamental file system design choice is whether it will allow file names to be case sensitive or not, and modern file systems are diverse in their selection.  A {\em case-sensitive file system} is one that allows the definition of multiple files whose names differ only in their case, such as {\tt Foo.c} and {\tt foo.c}. In a {\em case-insensitive file system}, only one file can be defined whose names differ only in their case.  Historically, UNIX file systems are case sensitive, whereas Windows file systems are case insensitive.  Further, case-insensitive file systems may be either case preserving (e.g. Apple File System (APFS), NTFS, etc.) or not (FAT), where a {\em case-preserving file system} preserves the case chosen (i.e., either {\tt Foo.c} or {\tt foo.c}), rather than converting all names to one case choice (e.g., all lowercase).  Importantly, while choices in case sensitivity for a single file system may appear to be arbitrary or aesthetically driven, the precise semantics of interactions between two file systems with different case sensitivities can range from subtle to ill-defined, with associated consequences.


Practitioners have long had concerns about the implications of leaving case sensitivity as an open design choice~\cite{linus-email}
Historically, these concerns were not considered as pressing when file systems were associated with their respective operating systems and associated singular assumptions about case.
However, \textit{individual systems now frequently support a mixture of case-sensitive and case-insensitive file systems}, creating opportunities for files to be moved between file systems with different cases and file identifier encodings.  More troublingly, {\em several file systems now support allowing the choice of case for individual directories}~\cite{ext4-support}, complicating file operations by having multiple case and encoding semantics within the same file system.  

Security risks related to this design choice therefore appear to be increasing. First, the Windows Subsystem for Linux~\cite{wsl} (WSL) integrates Linux and Windows platforms leading to expectations that files may be routinely copied from Linux (i.e., case-sensitive) to Windows (i.e., case-insensitive) file systems.
Second, Linux {\tt ext4} now supports case-sensitive and case-insensitive naming in the same partition, configurable per directory~\cite{ext4-support,lwn-icase-ext4}.  Linus Torvalds expressed concerns about adding such support to {\tt ext4}~\cite{linus-email}, stating that such features often cause "actual and very subtle security issues."  

Indeed, security issues caused by moving files from case-sensitive to case-insensitive file systems are starting to appear.  For example, the \git\ distributed version control system has suffered from multiple vulnerabilities (e.g., CVE-2014-9390, 
CVE-2021-21300), caused by how {\tt git} clones repositories from case-sensitive file systems to case-insensitive file systems.  
To exploit this, an adversary creates a repository in a case-sensitive file system with a directory whose name will {\em collide} (i.e., only differs in case) with a symbolic link (to another directory) added by \git\ when the repository is cloned to a case-insensitive file system.  The \textit{name collision} between the directory and the symbolic link enables adversaries to overwrite the scripts that \git\ executes.  Such attacks can alter both the target resource's content and/or its metadata, including its permission assignments.  


Researchers have long been aware of hazards that may occur during file system name resolution~\cite{mcphee74,bishop-dilger}, particularly that programmers must validate safe use of symbolic links and check for "squatted" files when creating a new files. Many defenses have been proposed~\cite{chari2010you,cowan2001raceguard,pf13eurosys,dean2004fixing,openwall,park2004rps,payer2012protecting,lhee2005detection,tsafrir2008portably,tsyrklevich2003dynamic,uppuluri2005preventing}. 
However, to the best of our knowledge, ours is the first work studying how case interplays cause name collisions that lead to incorrect, and in some cases, vulnerable behaviors. 
We show that utilities and applications currently do not recognize unsafe use of case-insensitive file systems, leading to these problems.  In particular, we study the behavior of six utilities used to copy files, which is how case-sensitive and case-insensitive file systems can interact. We find a wide variety of responses to name collisions, including many that overwrite existing data and change file permissions silently.  In addition, we examine three case studies in using such utilities that result in unsafe and sometimes exploitable behaviors.  
This paper demonstrates the potential implications of the name collision problem, focusing on Linux and its supported file systems, thereby motivating both more and broader (e.g., other OS-FS combination) investigations.

This paper makes the following contributions:

\begin{itemize}
    \vspace{-6pt}\item We examine the security and correctness implications of {\em name collisions}, when two distinct file system resources with two distinct names map to to a single name, due to file system case sensitivity and/or encoding mismatches.\vspace{-6pt}
    \item We develop an automated method to test common Linux utilities for unsafe reactions to name collisions, finding a wide variety of responses, many of which are unsafe and possibly exploitable.
    \vspace{-4pt}\item We perform detailed case studies of the impact of name collisions on three programs {\tt dpkg}, {\tt rsync}, and Apache, showing how they operate incorrectly in the face of name collisions and how they would be exploited when deployed on case-insensitive directories.\vspace{-4pt}
\end{itemize}

\section{Background: From Cases to Collisions}
\label{sec:background}

\label{subsec:icase}
Beyond traditional, i.e. operating-system-entailed, decisions made with respect to case sensitivity, even Linux files systems now represent a surprising diversity of case sensitivity decisions.
In particular, the desire to support some non-native applications, such as WINE and Samba from Windows systems, has motivated Linux file systems to support the case-insensitive file naming used in these non-native file systems. 

The ability to create case-insensitive file systems has long been possible in some Linux file systems, such as ZFS, JFS, and ciopfs.  However, these options are applied to the entire filesystem, rather than just the relevant directories for individual applications. 
In 2019, Linux kernel version 5.2 added support for per-directory case-insensitivity to ext4\cite{ext4-support,lwn-icase-ext4}. Later in 2019, similar support was added to the Flash-Friendly File System (F2FS) in Linux kernel version 5.4~\cite{f2fs-doc,f2fs-patch}.   For case-insensitive directories, these  file systems are case-preserving in nature.  

\vspace{-4pt}\subsection{Motivations for Increasing Case Diversity}
\noindent\textbf{Samba:}~~Samba~\cite{samba} implements the Common Internet File System (CIFS) protocol which allows for sharing file systems over a network. Its primary use is sharing files with Windows clients that expect a case-insensitive file system. Hence, Samba implements user-space case-insensitive lookups even if the underlying file system is case-sensitive. Furthermore, it allows turning on/off case-sensitivity and case-preservation on a per-mount basis~\cite{smb.conf}. Note that this feature only works for non-Windows clients, which means that the actual file system can contain files differing only in case. This can lead to unexpected behaviors where Samba will choose to show only a subset of files. Deleting files which have collisions will now show the alternate versions, thereby giving rise to inconsistent behavior from the end user's perspective.


Samba's requirement of case-insensitive matching, which is done in user-space, incurs a huge performance overhead~\cite{lwn-netfs} thereby motivating the support for case-insensitivity in the ext4 file system~\cite{lwn-icase-ext4,lwn-icase1,lwn-icase2}. Other programs/systems such as Wine~\cite{wine}, Network File System (NFS), SteamOS~\cite{steamos,steam-macos} and Android~\cite{lwn-android,xda-android} would also benefit from in-kernel case-insensitivity support.

\noindent\textbf{ext4:}~~For ext4, the idea is that the filesystem at large can be configured to be "casefolding," which permits the mixing of case-sensitive and case-insensitive directories in the same file system.  When creating an ext4 file system, the \emph{casefold} option is applied, e.g., {\tt mkfs -t ext4 -O casefold /dev/sda}.  Setting the {\tt +F} inode attribute on an empty directory makes it case-insensitive, e.g., {\tt mkdir foo; chattr +F foo}. Note that case-insensitive directories can contain case-sensitive directories. This means that for a given path, {\tt /foo/bar/bin/baz}, any of {\tt foo}, {\tt bar} and {\tt bin} can either be case-sensitive or case-insensitive.



\noindent\textbf{tmpfs:}~~{\tt tmpfs} recently added case-insensitivity support~\cite{lwn-case-tmpfs}. The use cases are similar to that of ext4 with the addition of supporting sandboxing and container tools such as Flatpak.


\subsection{Name Collisions} \label{sec:background_collision}

A {\em name collision} occurs when a file system maps two distinct names of two distinct resources to the same name.
Name collisions can cause problems to occur if the names of distinct resources \emph{collide} when those resources are replicated to a target directory that does not provide a 1:1 mapping for all replicated objects.  Suppose one directory has two files with distinct names in that file system. Should those files be copied to a second directory in which the two file names collide (i.e., are mapped to the same name), then only one file will be created, which may be either of the original files or an unpredictable combination of the two files' content and metadata.
Variation in case sensitivity between two file systems is a common origin of collisions, but diversity in other encoding properties, such as character choice (e.g., FAT does not support $``$, $:$, $*$, etc.~\footnote{\url{http://elm-chan.org/fsw/ff/doc/filename.html}}) and canonicalization processes, can lead to the same effect.   For example, NTFS uses UTF-16 while APFS (macOS) and ext4 (Linux) use UTF-8 and older file systems can use other encoding schemes, such as iso8859-1.

Modern file systems perform the canonicalization of names
using a technique called \emph{case folding}\cite{define-folding}.
Unlike traditional techniques, case folding uses lookup tables to transform each character of the filename to a pre-determined case.
Furthermore, in Unicode several different characters (or code points) can be used to represent the same abstract character. Hence, a normalization step is needed after performing case folding to ensure that the binary representations match.

Unfortunately, \textbf{\textit{these case folding rules can differ across file systems}}. Additionally, the \emph{locale} (or language) also influences the case folding rules.
Due to such differences, `temp\_200K' (where K = Kelvin Sign, U+212A in UTF-8) and `temp\_200k' are considered identical on NTFS and APFS, but on ZFS these filenames are considered different when using case-insensitive lookups. As a result, when two files of these names are copied from a ZFS file system to an NTFS file system, they will collide and only one filename and only one file will be created. Similarly, consider the filenames \f{floß}, \f{FLOSS} and \f{floss}. All can coexist on a case-sensitive file system supporting reasonable character encodings, but, since case-folding for both \f{floß} and \f{FLOSS} is \f{floss}, attempting to move these files to a case-insensitive system may only preserve one of the original triple.

Modern encoding schemes such as UTF-8 have support for non-English characters which require case folding to perform case-insensitive matching. This only increases the number of case-insensitive matches, making the problem of name collisions even worse. For clarity and conciseness, we will use examples of ASCII-based, case-insensitive matching throughout the rest of the paper.


\label{subsec:taxonomy}

\begin{figure}
\begin{center}
\resizebox{\linewidth}{!}{
\begin{forest}
[Name Confusion (NC)
	[Alias
    	[Symlink]
    	[Hardlink]
		[Bind mount]
	]
	[Squat
	    [File]
	    [Other]
	]
	[{\bf Collision}
		[{\bf Case}]
		[{\bf Encoding}]
	]
]
\end{forest}
}
\vspace{-0.26in}
\caption{Taxonomy of {\em name confusion vulnerabilities}  divided into {\em alias} (i.e., multiple names for a resource), {\em collision} (i.e., multiple resources for a name), and {squat} (temporal ambiguities in names vs. resources) classes}\vspace{-16pt}
\label{fig:tax}
\end{center}
\end{figure}
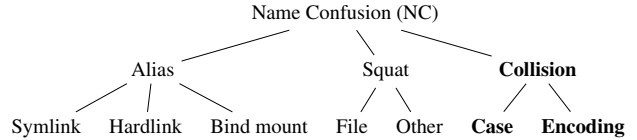

We propose a taxonomy for \emph{name confusions}, shown in \autoref{fig:tax}, that captures the types of incorrect program behaviors that may stem from the ambiguous uses of names for file system resources.
\textit{Name collisions} are a subset of this broader class. Name confusions may be caused by three reasons: (1) because multiple names may refer to the same resource (i.e., {\em aliasing}); (2) because an adversary may create a resource of that name before the victim (i.e., {\em squat}); and (3) because the multiple resources may be associated with the same name (i.e., {\em collisions}). Of these, however, name collisions are the least explored for their correctness and security implications. As Linux is adding more support for case-insensitivity, it is crucial to understand the pitfalls and problems such functionality may incur. This work aims to study these issues.


\section{From Collisions to Calamities}
\label{sec:motivation}

Name collisions can impair system functionality by modifying the content and/or metadata of files and directories in unexpected ways. Some name collisions have already led to security vulnerabilities~\cite{gitcve-fix}.  In this section, we define the conditions in which a name collision occurs, the conditions under which such a collision may be exploitable by an adversary, and describe a known vulnerability that is caused by a name collision.

\subsection{Causes of Name Collisions}
\label{subsec:model}

A process may cause a name collision under the following conditions.

\begin{itemize}
    \vspace{-6pt}\item There exists a \emph{source resource} (e.g., file or directory) in a case-sensitive file system, whose name is \emph{source name}.

    \vspace{-4pt}\item The process uses a relocation operation to place the source resource in a {\em target directory}, where the target directory is a case-insensitive or case-preserving directory.  Examples of relocation operations include copy (e.g., {\tt cp}, {\tt rsync}, or an archive operation, such as {\tt tar} or {\tt unzip}) or move (e.g., {\tt mv}).

    \vspace{-4pt}\item The relocation operation produces a {\em destination name} from the source name for the name of the source resource when placed in the target directory.

    \vspace{-4pt}\item There is a {\em target resource} with a {\em target name} whose name differs from the source name, but maps to the same name as the destination name does in the target directory (e.g., due to differences in case-folding rules between the source and target directory).  

    \vspace{-4pt}\item If the process is authorized to modify the target resource, the process's relocation operation results in a name collision between the target and source resources.  
    \vspace{-4pt}\item If the relocation operation proceeds despite the name collision, then the target resource's content and/or its metadata may be modified using the source resource content and/or metadata.
\end{itemize}\vspace{-4pt}

When these conditions are met, a name collision occurs such that the target resource in the target directory will be modified using the source resource.  In most cases, modifying a target resource using a source resource of a different name is an unexpected result.  We test how common Linux utilities react to name collisions and examine case studies where name collisions cause incorrect operation.  

Given the above conditions, there are several clear scenarios where the movement of files involving the following types of file systems (following the categorization in \autoref{sec:background_collision}) could result in name collisions:

\begin{itemize}
    \vspace{-6pt}\item Case-sensitive and case-insensitive file systems.
    \vspace{-4pt}\item Two distinct case-insensitive file systems with different case folding rules, e.g. ZFS to NTFS, etc.~\footnote{ASCII encoding contains a strict subset of characters representable under the UTF-8 encoding scheme. Hence, without loss of generality, we only consider ASCII encoded names for the examples in this paper.} 
    \vspace{-4pt}\item Two file system whose locales are different but they still use the same file system format (such as ext4).
    \vspace{-4pt}\item A single file system that supports per-directory case-insensitivity, e.g. ext4.\vspace{-2pt} 
\end{itemize} 

Clearly, name collisions may impact system functionality by causing collateral damage to resources supposedly unrelated to the operation, even removing the target resource entirely.  In addition, name collisions may be used to exploit the process performing the relocation operation in a version of a {\em confused deputy attack}~\cite{confused-deputy}.  An adversary only requires write access to the source directory to produce source names that may lead to name collisions to perform an attack.  We note that adversaries require fewer permissions to perform attacks using name collisions than other name confusion classes, which require write access to a directory used in name resolution of the target resource~\cite{sting12usenix}. Thus, remote attacks using file system archives, such as tarballs and zip files, as well as file repositories, such as GitHub, can be the sources of attacks.  

In practice, to perform a successful attack using a name collision, the victim process has to help the adversary in two ways.  First, the victim process has to use the source resource in a relocation operation planted by an adversary as described above.  
In addition to archives, other activities, such as backups, may provide opportunities for exploitation of name collisions.  In addition, ad hoc user actions copying files, e.g., from Linux to Windows in the Windows Subsystem for Linux, may result in unexpected and exploitable collisions.   Second, the target directory of the relocation operation has to be predictable by the adversary to enable them to produce a source name that leads to a colliding destination name.  Archives make this task much easier because the archive itself may be crafted to provide the target resource that is exploited by creating a collision with another archive file.  A recent vulnerability in the {\tt git} distributed revision control system demonstrates exactly this, as described below.

\subsection{An Example Collision Vulnerability}
\label{subsec:git}

Security vulnerabilities due to filename collisions across different file systems have been demonstrated in the wild. Consider a recent vulnerability in the \git\ distributed version control system (CVE-2021-21300).  
This vulnerability results in remote code execution after cloning a maliciously crafted repository created on a case-sensitive file system to a case-insensitive file system. 

\begin{figure}[ht]
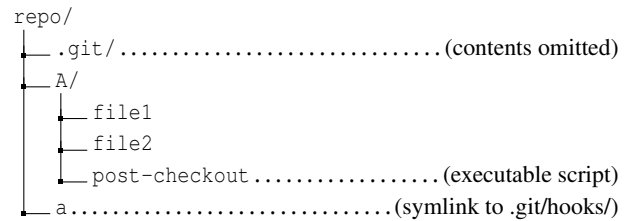

\vspace{-4pt}
\small
\dirtree{%
 .1 repo/.
 .2 .git/\DTcomment{(contents omitted)}.
 .2 A/.
 .3 file1.
 .3 file2.
 .3 post-checkout\DTcomment{(executable script)}.
 .2 a\DTcomment{(symlink to .git/hooks/)}.
}

\vspace{-6pt}\caption{Example for Git CVE-2021-21300}
\label{fig:git-example-motivation}\vspace{-6pt}
\end{figure}

\autoref{fig:git-example-motivation} depicts the maliciously crafted repository structure. 
Note that this directory structure works correctly on a case-sensitive file system.  However, on case-insensitive file systems, the presence of the `a' (small) and `A' (capital) directories creates a collision that exposes a vulnerability.
This collision results in a vulnerability when using \git's out-of-order checkout machinery.  Git Large File Storage (LFS) uses out-of-order checkouts for downloading binaries in the background.  Say the repository creator (adversary) marks `A/post-checkout' for an out-of-order checkout.  When a user clones this repository to a case-insensitive file system (e.g., NTFS), \git\ performs a sequence of operations that: (1) replaces `A' with the symbolic link `a' and (2) writes the script file `A/post-checkout' to `.git/hooks/post-checkout' due to the symbolic link `a'.  After the files are downloaded, \git\ runs the script `.git/hooks/post-checkout' that the adversary provided, which is obviously undesirable.  

In this case, a maliciously crafted \git\ repository can be designed to provide a target resource of the symbolic link `a', which when collided by `A' in resolving the source resource  `A/post-checkout' redirects the operation to a directory chosen by the adversary using the symbolic link.

\subsection{The State of Defenses for Name Confusions}
\label{subsec:relwork}

Currently, operating systems provide no innate defenses to prevent 
name collisions, leaving the challenge to programmers. 
However, researchers have studied problems due to other types of name confusions extensively, proposing a variety of defenses~\cite{chari2010you,cowan2001raceguard,dean2004fixing,openwall,park2004rps,payer2012protecting,lhee2005detection,tsafrir2008portably,tsyrklevich2003dynamic,uppuluri2005preventing,pu2006methodical}.  
However, researchers have shown that comprehensive program defenses are expensive~\cite{pf13eurosys} and that system-only defenses will always be prone to some false positives~\cite{cai09oakland}.  Leveraging limited program information~\cite{jigsaw14usenix,safe_open} still results in some false positives.   

As a result, library commands for opening files have been extended in a variety of ways to prevent name confusions from occurring.  The {\tt open} command has been extended with flags to detect file squats (i.e., {\tt O\_EXCL|O\_CREAT} to detect the presence of an existing file during file creation) and prevent unexpected use of aliases (i.e., {\tt O\_NOFOLLOW} to prevent following symbolic links).  However, the use of squats and aliases is desirable in some applications, despite their risks.  Thus, the {\tt openat} command has been added to enable programmers to avoid TOCTTOU attacks, by opening a file from a validated directory (i.e., file descriptor to the directory of the desired file).  The successful use of {\tt openat} requires the programmer to check for unwanted squats or aliases themselves.  An alternative is proposed by the {\tt openat2} command, which instead controls {\em how} files may be opened, such as requiring all file components accessed to be descendants of the directory from which the operation originates.  {\tt openat} and {\tt openat2} limit the attack surface of squat and alias attacks, but do not eliminate them entirely, depending on the programmer's additional actions to check for TOCTTOU attacks and configure the commands.  

At present, the above commands make no effort to help programmers address name collisions.  As a result, utilities to perform copy and move operations and applications that may utilize file systems with multiple or mixed (e.g., ext4 and F2FS) case sensitivities or encodings may not detect and resolve name collisions correctly.  We will examine the possible defenses for name collisions in \autoref{sec:defense}.

\section{Overview}
\label{sec:overview}

In this paper, we aim to explore the impact that name collisions may have on file system security.  To do this, we propose to examine three research questions.

{\bf RQ1}: {\em How do applications invoke utilities that may allow unsafe name collisions?}  In \autoref{sec:tooling}, we examine Linux packages to determine the most common options that applications employ for the utilities used to perform copies.  We examine how frequently application packages use utilities in copy operations by scanning their scripts for such operations, as shown in Table~\ref{tab:prevalence}.


{\bf RQ2}: {\em When do the utilities for performing copy operations allow unsafe name collisions?}  Recall that \autoref{subsec:model} defines the conditions under which an unsafe name collision may occur.  This research question asks whether the utilities that applications may use to perform copy operations (e.g., {\tt cp} and {\tt tar}) prevent unsafe effects when a name collisions occur.  For these utilities and the common options found in RQ1, we examine a variety of name collision scenarios to determine whether the utilities allow name collisions and their unsafe effects to occur as shown in \autoref{tab:tooling}.

{\bf RQ3}: {\em What correctness and security problems are caused by name collisions?}  In \autoref{sec:cases}, we examine three case studies where we show how name collisions cause programs to behave incorrectly.  In particular, we show concretely how applications can be vulnerable to name collisions when target resources are deployed on case-insensitive or case-preserving file systems.  

{\em Impacts:} A preview of our result is that: (1) many applications rely on these utilities to copy file system resources and repositories/archives; (2) the utilities used to copy file system resources and repositories/archives often allow unsafe name collisions, although the specific responses vary in ad hoc ways; and (3) applications currently lack defenses against name collisions, which can lead to incorrect operation and exploitable vulnerabilities.

\section{Testing for Name Collisions}
\label{sec:design}

This section details an automated tool for testing the responses of common Linux utilities used for relocation operations to name collisions.  As described in \autoref{subsec:model}, a name collision is caused by creating a source name that will be converted to a destination name by the relocation operation that is equal to a target name in the target directory of the operation.  Thus, our aim is develop a method to automate the generation of source resources with names that will lead to name collisions when relocated to case-insensitive targets and identify when operations allow the name collision to occur, detecting the effects of those operations. 

\subsection{Test Case Generation}
\label{subsec:case_gen}

The individual test cases are generated to test file system resources of various types, including regular files, directories, symbolic links (to files and directories), hard links, pipes, and devices.  In addition, we have found that creating collisions in non-trivial directory structures may also lead to incorrect behaviors.  \autoref{fig:test-example} shows an example test case where the directories as well as their contents result in a collision when transferred to a case-insensitive file system.  As a result, we aim to generate test cases that result in name collisions at different depths of the directory being copied, as evidenced by the collision between directory names at depth 2 (i.e., "dir" and "DIR") and the impact on colliding resources of different types (i.e., a regular file "foo" and a pipe "foo").  

\begin{figure}[ht]
\newcommand{\type}[1]{\textcolor{red}{#1}}
\centering\small
\begin{minipage}[t]{0.2\columnwidth}
    \textbf{\textsc{Input}}
    \dirtree{%
     .1 src/.
     .2 dir/.
     .3 foo\type{*}.
     .2 DIR/.
     .3 foo\type{|}.
    }
\end{minipage}
\begin{minipage}{0.3\linewidth}
    \centering
    \tikz[x=1ex,y=1em]
        {\draw
            [->] (0,0)
            --   (5,0) node [fill=white] {\textsc{copy}}
            --   (10,0)
        }
\end{minipage}
\begin{minipage}[t]{0.3\linewidth}
    {\centering\bfseries \textsc{Effect}}
    \dirtree{%
     .1 target/.
     .2 dir/.
     .3 foo\type{*|}.
    }
\end{minipage}
\\\vspace{2ex}\footnotesize
Here, `\type{*}' means a regular file and `\type{|}' means file type is a named pipe.
\vspace{-4pt}\caption{An Example Test Case}
\label{fig:test-example}
\undef\type
\end{figure}

Since we are testing the behavior of various utilities that perform relocation operations, we can control the source and target names in creating test cases.  As a result, the choice of names is trivial.  We create source directories that contain both the target resource (i.e., a resource copied first from the source to the target) and the source resource (i.e., a resource copied later by the utility to collide with the target resource (i.e., now in the target directory).  This is similar to the way name collisions would occur when copying an archive or repository that causes a collision, as the {\tt git} vulnerability.  Since different utilities may process resources in different orders, we generate test cases with both orderings of resources that may cause collisions.   

\begin{figure*}
\centering
\newcommand{\highlight}[1]{\textcolor{red}{#1}}\small
\begin{tikzpicture}[
    x=1em,y=1ex,remember picture,
    tag/.style={color=blue,font={}},
]
    \node[anchor=north west] (use) at (0,0) {\tt
        \subnode{tool use}{\textbf{USE}}
        [msg=10960,\subnode{tool bin}{`cp'}.openat]
        \subnode{tool id}{00:39|2389}|
        /mnt/folding/dst/\highlight{ROOT} $\hookleftarrow$
    };

    \node[anchor=north west] (create) at (0,-3) {\tt
        \subnode{tool create}{\textbf{CREATE}}
        [\subnode{tool auditd}{msg=10957},`cp'.
        \subnode{tool syscall}{openat}]
        00:39|2389|
        \subnode{tool path}{/mnt/folding/dst/\highlight{root}}
    };

    \draw [decorate,decoration={brace}]
        (tool id.north west) --  (tool id.north east)
        node[tag,pos=0.5,above=0.5ex] {device | inode};

    \draw [decorate,decoration={brace}]
        (tool bin.north west) --  (tool bin.north east)
        node[tag,pos=0.5,above=0.5ex] {program};

    \draw [decorate,decoration={brace,mirror}]
        (tool path.south west) --  (tool path.south east)
        node[tag,pos=0.5,below=0.5ex] {accessed path};

    \draw [decorate,decoration={brace,mirror}]
        (tool syscall.south west) --  (tool syscall.south east)
        node[tag,pos=0.5,below=0.5ex] {syscall};

    \draw [decorate,decoration={brace,mirror}]
        (tool auditd.south west) --  (tool auditd.south east)
        node[tag,pos=0.5,below=0.5ex] {auditd id};

    \node[anchor=west,tag] (op) at (2,-12) {operation};
    \draw [->] (op) to (tool create.south);
    \draw [->] (op) to[out=150,in=190] (tool use.west);

    \begin{scope}[on background layer]
        \path [fill=gray!5,rounded corners]
            (use.north west) rectangle (create.south east);
    \end{scope}
\end{tikzpicture}
\vspace{-12pt}\caption{Example violation reported by name collision testing}\vspace{-8pt}
\label{fig:mon-tool}
\undef\highlight
\end{figure*}
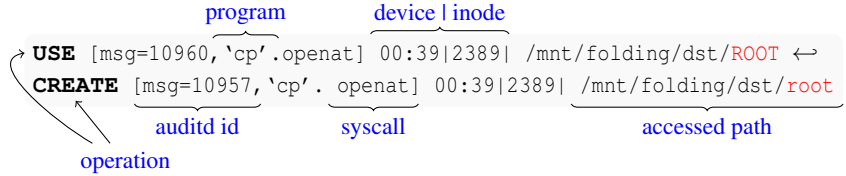

The only decisions then are what are the resource types of the source and target resources and where to place them in the source directory hierarchy to cause the desired collision to be created.  Symbolic links, pipes, and devices only create interesting behaviors when used as target resources.  For symbolic links, the unsafe effect is to follow the link to another file system resources, which only happens with the symbolic link as the target resource.  For pipes and devices, the unsafe effect is to send the source resource's content to the pipe or device, which also is only possible if these are target resources.  

As a result, the automated test generation produces test cases consisting of source and target resources of all combinations of potentially unsafe resource types and places these test cases at depth one and/or two of the file system hierarchy.  For {\tt rsync}, we specifically found an issue caused by a collision at depth two, without any collision at depth one, as detailed in Section~\ref{subsec:rsync}.  






\subsection{Detecting Collision Effects}
\label{sec:tracing}

The key idea is to record the file system operations sufficiently to detect that an unsafe name collision has occurred.  Since we design the test cases to create a name collision on a relocation operation, we want to detect when such an operation is a successful collision.  Then, we need to determine the impact of the operation to classify the effect according to one of the ten effect options defined in \autoref{sec:utility-experiment}.

We monitor file system operations using {\tt auditd} to detect successful collisions.  An example of a log indicating a collision is shown in \autoref{fig:mon-tool}.  In this example, a {\em create} operation creates a target resource named ``root''  using {\tt openat} command, but a later {\em use} operation to the same resource (i.e., same device-inode pair, see below) is associated with a name ``ROOT'',
which differs from the name used when the resource was created.  Note that although the target resource was created on a case-insensitive file system, multiple names may be used that are resolved to the same name.

We say that a collision is successful when we detect a {\em use} of a target resource with a different name than that used to create the target resource.  To detect such collisions, we first identify the file system operations that {\em create} a target resource, recording its combination of device and inode identifiers, which form a unique resource identifier and its pathname.  In \autoref{fig:mon-tool}, the name component ``root'' will be important to detecting the collision.
We then capture all the file system operations that {\em use} the target resource.  In general, if a file pathname in a successful {\em use} operation on a target resource does not match the file pathname of a {\em create} for the same resource, then we record a positive (i.e., a collision).  In \autoref{fig:mon-tool}, the pathname of the {\em use} operation differs between ``root'' and ``ROOT'', indicating a name collision.

We also record a positive when a {\em use} operation deletes and replaces a resource from a prior {\em create} operation, as some collisions may cause the target resource to be deleted and the source resource to replace it.  We validate that there is a {\em create} operation for the colliding destination name to verify the cause of the deletion is a collision.



To detect the effect of a name collision, we examine the resulting resource that now maps to the target name.  We compare the source resource and target resource content and metadata to the resultant resource to determine whose content and/or metadata (i.e., source, target, or neither) the resource has.  For tests on directories and hardlinks, we examine the directories and the resultant directory entries.

\section{Name Collisions on Linux Copy Utilities}
\label{sec:tooling}

\begin{table*}[ht]
  \caption{Prevalence of copy utilities}
  \label{tab:prevalence}

  \centering
  \resizebox{\textwidth}{!}{
  \begin{tabular}{c|c|c|c|c}
    \toprule
    \texttt{tar} & \texttt{zip} & \texttt{cp} & \texttt{cp*} & \texttt{rsync} \\
    \midrule

    \begin{tabular}{cl}
        10 & mc \\
        8  & perl-modules \\
        7  & libkf5libkleo-data\\
        6  & pluma \\
        6  & mc-data \\
           & \ldots \\
        \midrule
        107 & TOTAL
    \end{tabular} &

    \begin{tabular}{cl}
        21 & texlive-plain-generic \\
        15 & aspell \\
        11 & libarchive-zip-perl \\
        7  & texlive-latex-recommended \\
        5  & texlive-pictures \\
           & \ldots \\
        \midrule
        69 & TOTAL
    \end{tabular} &
    
    \begin{tabular}{cl}
        78 & hplip-data \\
        32 & dkms \\
        22 & libltdl-dev \\
        20 & autoconf \\
        18 & ucf \\
           & \ldots \\
        \midrule
        538 & TOTAL
    \end{tabular} &

    \begin{tabular}{cl}
        12 & dkms \\
        2  & udev \\
        2  & debian-reference-it \\
        2  & debian-reference-es \\
        1  & zsh-common \\
           & \ldots \\
        \midrule
        25 & TOTAL
    \end{tabular} &

    \begin{tabular}{cl}
        28 & mariadb-server \\
        5  & duplicity \\
        4  & texlive-pictures \\
        2  & vim-runtime \\
        1  & rsync \\
           & \ldots \\
        \midrule
        42 & TOTAL
    \end{tabular}\\
    \bottomrule
  \end{tabular}
  }

  \raggedright\footnotesize\vspace{1ex}
  We calculate the number of times that each command (tar, zip, etc.) is used inside scripts from various packages. We investigate 4752 .deb packages from the installation disk (DVD \#1) of \emph{Debian 11.2.0}. Only the top-five packages are shown (entries are sorted in descending order for each command).\vspace{-8pt}
\end{table*}

%
%
%


In this section, we examine how common Linux utilities that applications use to copy files\footnote{We focus on copy operations.  The impact on move operations is similar because (unless both target and source are on the same file system) it simply performs a copy first and then deletes the source.}
from one part of the file system to another react when the copy operation causes a name collision in a case-insensitive directory. To quantify the ubiquity of these utilities, we survey their use by packages on Debian 11.2.0. We retrieve all packages from the Debian installation DVD and count the number of times the copy utilities are used inside the packages' scripts. The results are summarized in \autoref{tab:prevalence}.
Note that the listed uses of these utilities are lower bounds because we do not parse executable binaries. Hence, we miss uses where the utilities are invoked via system calls such as {\tt system(...)}, {\tt execve(...)}, etc.


\subsection{Collecting Responses to Name Collisions}
\label{sec:utility-experiment}

  \newcommand{\YES}{\textcolor{red}{\bfseries\checked}}
  \newcommand{\DEL}{\textcolor{red}{$\times$}}
  \newcommand{\OW}{\textcolor{red}{$+$}}
  \newcommand{\A}{$A$}
  \newcommand{\E}{$E$}
  \newcommand{\R}{$R$}
  \newcommand{\M}{\textcolor{blue}{$\neq$}}
  \newcommand{\U}{\textcolor{gray}{$-$}}
  \newcommand{\T}{\textcolor{gray}{$T$}}
  \newcommand{\C}{\textcolor{red}{$C$}}
  \newcommand{\INF}{$\infty$}

\newcommand{\undefstrategies}{
  \undef\YES
  \undef\DEL
  \undef\OW
  \undef\A
  \undef\E
  \undef\R
  \undef\M
  \undef\U
  \undef\T
  \undef\C
  \undef\INF
}

\begin{table*}[ht]
  \caption{Name Collision Responses for Popular Linux Utilities}
  \label{tab:tooling}

  \centering
  \small
  \begin{tabular}{ll |c|c|c:c|c|c}
    \toprule
    \multicolumn{2}{c}{Name Collision between} \\
    \rule{0pt}{0ex} 

    Target Type
    & Source Type
    & \texttt{tar}
    & \texttt{zip}
    & \texttt{cp}
    & \texttt{cp*}
    & \texttt{rsync}
    & Dropbox
    \\
    \midrule



    file & file
        & \DEL & \A   & \E & \OW\M  & \OW\M  & \R  \\
    symlink (to file) & file
        & \DEL & \A   & \E & \OW\T  & \OW\M   & \R  \\
    pipe/device & file
        & \DEL & \U   & \E & \OW    & \OW     & \U  \\
    hardlink & file
        & \DEL & \U   & \E & \OW\M  & \OW\M   & \U  \\
    hardlink & hardlink
        & \C\DEL& \U   & \E & \C\DEL& \C\OW\M & \U  \\
    
    \midrule
    directory & directory
        & \OW\M  & \OW\M  & \E & \OW\M  & \OW\M   & \R  \\
    symlink (to directory) & directory
        & \OW  & \INF & \E & \E   & \OW\T  & \R  \\ 
    \bottomrule
  \end{tabular}

  \vspace{1.5ex}
  \begin{minipage}{1.0\linewidth}
  This table shows results of copying files/directories from a case-sensitive to a case-insensitive file system. {\tt cp*} refers to {\tt cp} being used with shell completion. For e.g., {\tt `cp * /target'} which copies all items from the current directory to {\tt /target} directory.
  \newline
  \end{minipage}
{\small
  \begin{minipage}[t]{0.33\textwidth}
    \begin{description}
        \item[\DEL] Delete existing file and create new file\vspace{-4pt}
        \item[\OW]  Overwrite existing file. For directories, merge their contents.\vspace{-4pt}
	    \item[\M] Mismatch between content and metadata\vspace{-4pt}
    \end{description}
  \end{minipage}
  \hspace{0.2in}
  \begin{minipage}[t]{0.31\textwidth}
    \begin{description}
  	    \item[\A] Ask user to resolve the collision\vspace{-4pt}
  	    \item[\T] Traverse symlink\vspace{-4pt}
	    \item[\C] Corrupts non-colliding files\vspace{-4pt}
	    \item[\E] Deny operation and report error\vspace{-4pt}
    \end{description}
  \end{minipage}
  \hspace{-0.1in}
    \begin{minipage}[t]{0.32\textwidth}
    \begin{description}
	    \item[\INF] Program crashes, or hangs\vspace{-4pt}
    	\item[\U] Ignore unsupported file type (for hardlinks create regular file instead)\vspace{-4pt}
	    \item[\R] Rename colliding file/directory\vspace{-4pt}
    \end{description}
  \end{minipage}
}
\end{table*}

The name collision test cases and the responses of copy utilities are shown in \autoref{tab:tooling}.  
The `Target Type' column represents the resource type of the target resource that may be overwritten. The `Source Type' represents the resource type of the source that collides with the target. The rest of the columns represent individual utilities and their responses to name collisions between a source resource of the source type and a target resource of the target type. 

Below is a comprehensive list of the types of responses observed.  Only "Deny" and "Rename" prevent name collisions from causing unsafe and possibly exploitable behaviors, although both may block legitimate functionality in some cases.  "Ask the User" may result in an unsafe response if the user allows the target resource to be overwritten. Note that more than one response is possible for each test case.  

\begin{description}
    \vspace{-4pt}\item[Delete \& Recreate (\DEL)]  \emph{Delete} the target resource and \emph{create} a new resource based on the source resource.  The new resource's type, as well as its data and metadata, is determined by the source resource. The target resource is lost without any notification.\vspace{-4pt}
    
    
    
    \item[Overwrite (\OW)]  {\em Overwrite} the data and metadata of the target resource using the source resource. Unlike {\em Delete \& Recreate}, the name of the target resource is preserved. If file {\tt foo} is being overwritten with file {\tt FOO}, then the final file will be named {\tt foo} but will have the contents and metadata of file {\tt FOO}.\vspace{-4pt}

    \item[Corrupt (\C)] Contents of a resource that is not involved in name collision (i.e., not the target resource) is modified. For a more in-depth discussion, refer to \autoref{sec:hardlink-corruption}.\vspace{-4pt}

    \item[Metadata Mismatch (\M)] After a successful copy of a given source resource, some metadata, such as its name, UNIX permissions, user or group ID, extended attributes, or timestamp, remain from the target resource, creating a resource with a {\em mismatch} between the data (from the source) and the metadata (from the target).\vspace{-4pt}

    \item[Follow Symlink (\T)] Follow {\em symbolic links}, \textit{even when explicitly directed not to do so}.

    \item[Rename (\R)] The source name is {\em renamed} automatically to avoid creating a name collision, such as by appending a counter, resulting in a copy of the source resource in the target resource's directory with a non-colliding name.\vspace{-4pt}

    \item[Ask the User (\A)] To resolve a collision, {\em ask the user} to choose from a list of actions, such as to overwrite the target resource, skip copying the source resource, rename the target resource, abort, etc.

    Note that the user can still choose a response that results in adverse consequences. For instance, if the user chooses to overwrite the target, the target's data and metadata are modified using the source.\vspace{-4pt}

    \item[Deny (\E)] {\em Deny} the copy associated with a collision and report an error.\vspace{-4pt}

    \item[Crashes (\INF)] Collisions can result in the program hanging (e.g., going into an infinite loop) or {\em crashing}.\vspace{-4pt}

 

    \item[Unsupported file type (\U)] Does not support copying a resource if the source resource is of this file type.
    Note that if hardlinks are not recognized by a utility, then it simply creates a fresh copy of the underlying file.
\end{description}

The exact command-line flags used used to generate \autoref{tab:tooling} are listed in \autoref{sec:cmd-flags}.
To identify these flags, we analyzed 4,752 {\tt .deb} packages on  Debian 11.2.0's installation DVD. We found that the most commonly used flags enabled the following functionality.

\begin{itemize}
    \vspace{-4pt}\item Support recursively copying all directories.\vspace{-4pt}
    \item Support copying symbolic-links and hard-links as-is but {\em do not follow} them.\vspace{-4pt}
    \item Preserve metadata such as UNIX permissions, extended attributes (xattr), timestamps, and owner/group IDs (uid/gid).
\end{itemize}

Before examining the responses in Table~\ref{tab:tooling}, we briefly note some additional context for two of the columns. 

\paragraph{cp vs. cp*}
Both of these represent the same executable binary. The difference is in the way the command-line arguments are passed to the binary. Specifically, the format of specifying the source directories is different.

Consider that the source directory (to be copied) is {\tt foo}. For {\tt cp}, we will pass it as {\tt foo/} while for {\tt cp*} we will use {\tt foo}. Note the trailing {\tt /} is missing in the latter case. Just this difference significantly changes the behavior of {\tt cp} as noted in \autoref{tab:tooling}.

We use the {\tt cp*} method of invocation coupled with shell completion, e.g., `{\tt cp src/* /target}' where the shell replaces {\tt src/*} with each individual entry present inside {\tt src} sans the trailing {\tt /}. When testing the {\tt cp} method, we change the command to `{\tt cp src/ /target}'.

\paragraph{Dropbox}

Strictly speaking, \emph{Dropbox}\cite{dropbox} is not a copy utility but a popular file synchronization utility. It is intended to replicate entire directories across multiple machines and file systems.

We mention Dropbox to highlight its distinct response to handling {\em potential} name collisions.  Even when the underlying file system is case-sensitive, Dropbox treats it as {\em case-insensitive}. It proactively renames the files and directories to avoid name collisions that could occur if they were transferred to a case-insensitive file system. Note, however, that its renaming strategy is not even uniform across platforms: For example, the Dropbox application appends ``(Case Conflicts)'', ``(Case Conflicts 1)'', etc. to the file/directory names in case of a potential collision, whereas, when using their web-based interface, they append ``(1)'', ``(2)'', etc. instead.


\subsection{Unsafe Responses to Name Collisions}
\label{sec:tooling-security}

Several responses shown in \autoref{tab:tooling} demonstrate that utilities often allow unsafe responses to name collisions.
In this section, we examine some of the more concerning responses to show how utilities delegate responsibility for security against name collisions to the applications that invoke them.
For the examples in upcoming sections, {\tt src/} and {\tt target/} are on case-sensitive and case-insensitive file systems respectively.

%


\subsubsection{Silent data loss with {\em tar, cp* \& rsync}}

Name collisions involving files generally result in silent data loss. From \autoref{tab:tooling}, we can see that {\tt tar} deletes and recreates (\DEL) files when collisions occur. Hence, when there is a name collision between {\tt foo} and {\tt FOO}, only one of these files will remain in the target directory. The other file is permanently lost without any notification.

Similar to {\tt tar}, {\tt cp*} and {\tt rsync} also lose files silently. However, their behavior of overwriting (\OW) files results in other problems that are discussed later in this section. 

Unlike {\tt tar}, {\tt zip} and {\tt cp} will ask a user for next steps (\A) or report an error (\E) respectively. Hence, they are not prone to silently losing files.

\subsubsection{Merge directories with {\em tar, zip, rsync \& cp*}}

\begin{figure}[ht]
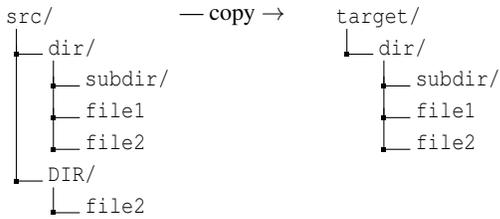
\vspace{-4pt}
\centering\small
\begin{minipage}[t]{0.3\linewidth}
    \dirtree{%
     .1 src/.
     .2 dir/.
     .3 subdir/.
     .3 file1.
     .3 file2.
     .2 DIR/.
     .3 file2.
    }
\end{minipage}
\begin{minipage}{0.2\linewidth}
    \centering
    --- copy $\rightarrow$
\end{minipage}
\begin{minipage}[t]{0.3\linewidth}
    \dirtree{%
     .1 target/.
     .2 dir/.
     .3 subdir/.
     .3 file1.
     .3 file2.
    }
\end{minipage}
\vspace{-4pt}\caption{Impact of merging directories}\vspace{-4pt}
\label{fig:merging-dirs}
\end{figure}

Name collisions involving two directories results in their contents (files, directories, etc. inside the directory) being merged. All of {\tt tar, zip, rsync,} and {\tt cp*} will silently merge directory contents without notifying the user. \autoref{fig:merging-dirs} highlights this issue using a directory listing.

In this example in \autoref{fig:merging-dirs}, the data of file {\tt file2} is overwritten by the content written last in the copy operation.  For example, if {\tt src/DIR}'s contents are written last, then its content for {\tt file2} is preserved and {\tt src/dir}'s is lost.

Furthermore, when the colliding directories have different UNIX permissions, a collision results in metadata mismatch (\M). With respect to \autoref{tab:tooling}, the UNIX permissions of the target resource are overwritten with permissions of the source  resource.

In \autoref{fig:merging-dirs}, consider {\tt src/dir/} with perms=700 and an adversary who creates {\tt src/DIR/} with perms=777. After a copy (using any of the above utilities), {\tt target/dir/} will have perms=777 effectively giving the adversary permission to the contents of the original {\tt src/dir/}.

\subsubsection{Stale names}

Whenever utilities resort to overwriting (\OW), we end up with stale file/directory names. For example, consider a name collision between a target resource {\tt foo} (file content: `bar') and a source resource {\tt FOO} (file content: `BAR'). After copying with {\tt rsync} or {\tt cp*}, we will end up with file {\tt foo} whose contents are `BAR'. 

The problem with such name collisions is that to the end user (or other programs), it will appear that {\tt foo} was successfully copied while in reality {\tt FOO} was copied. Just using the filename is not enough to discern which files were successfully copied. This is especially true for case-preserving file systems where the user has the expectation of the filenames being preserved. Hence, it is not unreasonable for the user to expect {\tt foo} should contain {\tt bar}.

\subsubsection{Symbolic link traversal at target}
Name collisions between {\em symlink (to file) and a regular file} results in {\tt cp*} following the symlink (\T) and overwriting (\OW) its target's contents with that of the regular source file.
With regards to \autoref{tab:tooling}, if the target resource is a symbolic link and the source resource is a file, then {\tt cp*} ends up following the symlink and writing data to the resource referenced by the symlink.

\begin{figure}[ht]
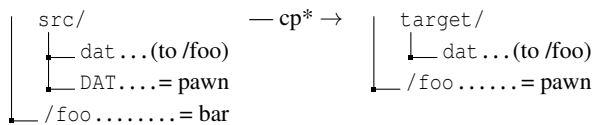

\centering\small
\begin{minipage}[t]{0.35\linewidth}
    \dirtree{%
     .1 src/.
     .2 dat\DTcomment{(to /foo)}.
     .2 DAT\DTcomment{= pawn}.
     .1 /foo\DTcomment{= bar}.
    }
\end{minipage}
\begin{minipage}{0.2\linewidth}
    \centering
    --- cp* $\rightarrow$
    
\end{minipage}
\begin{minipage}[t]{0.35\linewidth}
    \dirtree{%
     .1 target/.
     .2 dat\DTcomment{(to /foo)}.
     .1 /foo\DTcomment{= pawn}.
    }
\end{minipage}
\vspace{-4pt}\caption{Following symlink}\vspace{-4pt}
\label{fig:follow-symlink}
\end{figure}

\autoref{fig:follow-symlink} illustrates this case with an example. {\tt src/dat} is a symbolic link to {\tt /foo} and {\tt /foo} contains `bar'. Mallory (our adversary) does not have write access to {\tt /foo} but does have access to {\tt src/}. She creates {\tt src/DAT} which contains `pawn'.

Then the administrator starts the copy using: {\tt cp -a src/* target/}. At this point, {\tt cp} first creates the symlink {\tt target/dat}. Then it overwrites (\OW) this symlink with the contents of {\tt src/DAT}, effectively updating the file {\tt /foo}. After the copy has completed, {\tt /foo} contains `pawn'.

{\em cp*} has no command-line options to prevent traversal of symbolic links at the target. Only link traversal at the source can be turned off via command-line flags.

\subsubsection{The case of {\em hardlink -- hardlink} name collisions}
\label{sec:hardlink-corruption} 

During a copy when hardlinks (whose targets are different) collide, it can corrupt (\C) other non-colliding files and create spurious hardlinks. \autoref{tab:tooling} shows that this behavior is exhibited by {\tt tar, cp*}, and {\tt rsync}. An interesting observation is that, regardless of whether the utility's behavior is {\em Delete \& Recreate} (\DEL) or  {\em Overwrite} (\OW), this problem affects both.



\begin{figure}[ht]
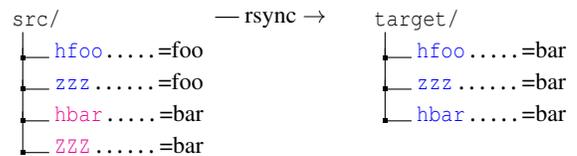

\newcommand{\one}[1]{\textcolor{blue}{#1}}
\newcommand{\two}[1]{\textcolor{magenta}{#1}}
\centering\small
\begin{minipage}[t]{0.35\linewidth}
    \dirtree{%
     .1 src/.
     .2 \one{hfoo}\DTcomment{=foo}.
     .2 \one{zzz}\DTcomment{=foo}.
     .2 \two{hbar}\DTcomment{=bar}.
     .2 \two{ZZZ}\DTcomment{=bar}.
    }
\end{minipage}
\begin{minipage}{0.2\linewidth}
    \centering
    --- rsync $\rightarrow$
    
\end{minipage}
\begin{minipage}[t]{0.35\linewidth}
    \dirtree{%
     .1 target/.
     .2 \one{hfoo}\DTcomment{=bar}.
     .2 \one{zzz}\DTcomment{=bar}.
     .2 \one{hbar}\DTcomment{=bar}.
    }
\end{minipage}
\vspace{-4pt}\caption{{\em hardlink -- hardlink} name collision}\vspace{-10pt}
\label{fig:hardlink-flaw}
\undef\one
\undef\two
\end{figure}

To understand this scenario, consider \autoref{fig:hardlink-flaw} that uses {\tt rsync} to perform the copy. The same color coding represents files that are hard-linked to each other. So {\tt src/hfoo} and {\tt src/zzz} are hard-linked, representing the same file. These files contain `foo'. Similarly, {\tt src/hbar} and {\tt src/ZZZ} are hard-linked and they contain `bar'.

After copying using {\tt rsync}, {\tt target/} contains three files that are all hard-linked to each other. Unlike the {\tt src/} directory, {\tt target/hfoo}, {\tt target/hbar}, {\tt target/zzz} are all hardlinks of each other and they contain `bar'.

Additionally, note that although the name collision happened between {\tt zzz} and {\tt ZZZ}, the contents of {\tt hfoo} were replaced. Even {\tt tar}, which deletes the old file and recreates it, exhibits this behavior.

The following order of operations undertaken by {\tt rsync} result in this behavior.


\begin{enumerate}
    \vspace{-4pt}\item Copy {\tt src/hbar} to {\tt target/hbar}. Now {\tt target/hbar} contains `bar'.\vspace{-4pt}
    
    \item Copy {\tt src/zzz} to {\tt target/zzz}. Now {\tt target/zzz} contains `foo'.\vspace{-4pt}
    
    \item In {\tt target/}, hardlink {\tt ZZZ} to {\tt hbar}. Due to NC, this effectively changes {\tt zzz} to be hard-linked to {\tt hbar}. Now {\tt target/zzz} contains `bar'.\vspace{-4pt}

    \item In {\tt target/}, hardlink {\tt hfoo} to {\tt zzz}. Now {\tt target/foo} contains `bar'. Additionally, all three files inside {\tt target/} are hard-linked to each other.\vspace{-4pt}
\end{enumerate}

The above copy is semantically different from the {\tt src/}. Specifically, name collision results in {\em distinct} sets of files getting hard-linked with each other at the {\tt target/}.

\section{Case Studies}
\label{sec:cases}

In this section, we examine case studies where name collisions cause unsafe behaviors, some of which are exploitable.  


\subsection{\texttt{dpkg} Package Manager}
\label{subsec:dpkg}

{\tt dpkg} is the package manager on Debian OS and its derivatives such as Ubuntu. {\tt dpkg} packages are compressed tarballs with extension {\tt .deb}.  When {\tt dpkg} processes a package, it tracks all files it creates during package installations in a database. Before installing a new package, {\tt dpkg} leverages this database to ensure that any files of previously installed packages will not by overwritten by this new package thereby preventing potentially malformed packages from corrupting the system.

On the other hand, we have observed that {\tt dpkg} will allow a package installation to replace any file whose name is not in its database, even privileged user files.  Thus, as long as a file in a package has a filename that does not match the filename of another package's file, {\tt dpkg} will install the file, silently replacing any existing file.

However, regardless of the underlying file system, the above database is matched in a \emph{case-sensitive} manner. This allows new packages to \emph{replace files} of previously installed packages via name collisions effectively circumventing the safeguards in {\tt dpkg}.  


In addition, and perhaps even more seriously, dpkg may allow an adversary to replace a package's customized config file with the default, reverting important changes.  {\tt deb} packages can mark certain files as configuration (or config) files. During package upgrades, if {\tt dpkg} spots modifications to these config files then it prompts the user to review the changes.

However, the config files are also matched in a case-sensitive manner. Under name collisions, {\tt dpkg} will just replace the original package's config file with the config file of the new package. For services, such as {\tt sshd}, {\tt httpd}, etc., config files are critical to their security, so such overwrites can potentially make the system vulnerable..

\paragraph{Reporting}
We have reported these issues to the maintainers of {\tt dpkg}.  The maintainers of {\tt dpkg} have since updated their package documentation\cite{dpkg-doc-diff} to warn end user communities not to use {\tt dpkg} where targets may be case-insensitive (i.e. specific directories, or entire file systems).

During our discussions, we analyzed 74,688 packages and found 12,237 filenames from those packages would collide if a case-insensitive file system were used, breaking multiple packages that contain these files. The name collision problem is fundamentally entrenched into the way {\tt dpkg} is implemented because it reasons about \emph{names} without involving the underlying file system(s).

\subsection{Rsync}
\label{subsec:rsync}

%

{\tt rsync} demonstrates vulnerable behavior when processing name collisions involving \emph{directories}.   During copy, the default behavior of {\tt rsync} is to simply recreate the symbolic links present at source.  However, when colliding directories contain sub-directories and symbolic links with the same name, the collision causes {\tt rsync} to suffer from link traversal\footnote{In this case, the name collision makes the alias exploitable, again combining name confusions.}.

\begin{figure}[ht]
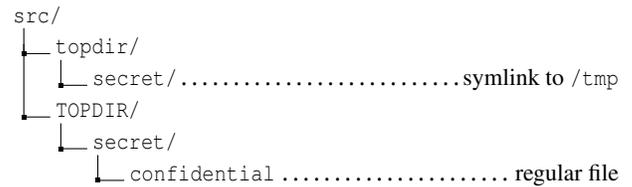
\vspace{-4pt}
\small
\dirtree{%
 .1 src/.
 .2 topdir/.
 .3 secret/\DTcomment{symlink to {\tt /tmp}}.
 .2 TOPDIR/.
 .3 secret/.
 .4 confidential\DTcomment{regular file}.
}
\vspace{-4pt}\caption{Case-sensitive source that {\tt rsync} is copying}
\label{fig:rsync-sourcefs}
\end{figure}

Consider the source directory listed in \autoref{fig:rsync-sourcefs}. Here, the directories {\tt topdir/} and {\tt TOPDIR/} only differ in case. So when copying to a case-insensitive file system, {\tt rsync} will encounter a name collision.

We use the following command to perform the copy:
\begin{center}
\vspace{-2pt}{\tt rsync -a src/ dst/}\vspace{-2pt}
\end{center}

\begin{tabular}{ c p{0.8\linewidth} }
    where,\\
    {\tt -a} & recursively copy directories, preserve symlinks, timestamps, DAC \\
    {\tt src/} & is case-sensitive \\
    {\tt dst/} & is case-insensitive \\
\end{tabular}
\newline

\begin{figure}[ht]
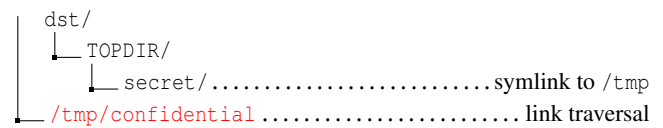

\small
\dirtree{%
 .1 dst/.
 .2 TOPDIR/.
 .3 secret/\DTcomment{symlink to {\tt /tmp}}.
 .1 \textcolor{red}{/tmp/confidential}\DTcomment{link traversal}.
}
\vspace{-4pt}\caption{After copying to case-insensitive destination}\vspace{-8pt}
\label{fig:rsync-targetfs}
\end{figure}

After the copy is completed, the newly created files are shown in \autoref{fig:rsync-targetfs}. Note that the file named \textcolor{red}{\tt confidential} ends up in {\tt /tmp}.

{\tt rsync} has created the \textcolor{red}{\tt /tmp/confidential} file by following the symbolic link {\tt dst/TOPDIR/secret}.

Below, we describe how this situation can be exploited.
Consider an adversary who wants to access a confidential file in {\tt TOPDIR/} to which she lacks any access.  However, she knows that {\tt TOPDIR/} is processed by a backup operation using {\tt rsync}. If she can create a sibling directory {\tt topdir/}, to which she will have read-write access, she can direct {\tt rsync} to write the confidential file (inside {\tt TOPDIR/}) to any directory of her choosing by creating a symbolic link inside {\tt topdir/} to that directory.

\paragraph{Reporting}
We reported this issue to the {\tt rsync} maintainers, and they told us that user's should not use {\tt rsync} with non-case honoring file systems.
However, we have concerns about the user community following such a recommendation in this case, since {\tt rsync} is often used by individuals.

In the course of these discussions, we learned the cause of the incorrect behavior.  {\tt rsync} assumes a one-to-one mapping of directories between source and target file systems. When a name collision results in two source directories being mapped to a single directory in the target, {\tt rsync} can be tricked into incorrectly predicting the target file type. In the presented scenario, a symbolic link {\tt src/topdir/secret} (to a directory) is incorrectly inferred to be a regular directory {\tt src/TOPDIR/secret}.

{\tt rsync} uses the {\tt O\_NOFOLLOW} flag with {\tt open()} to prevent link traversal and uses {\tt openat()/openat2()} to contain link traversals within a directory hierarchy, but it fails correctly use them because it assumes the symbolic link {\tt topdir/secret} is a directory.  

\subsection{Apache httpd}
\label{subsec:httpd}



Security of certain applications relies on the security parameters of the underlying file system. One such application is Apache's {\tt httpd}. It allows access to the underlying file system via the {\tt HTTP} protocol, relying on the UNIX Discretionary Access Control (DAC) permissions\footnote{If the system supports Mandatory Access Control (MAC), then DAC is used in conjunction with MAC.} to mediate the access. For example, files are accessible over {\tt HTTP} only if: (i) its UNIX group is {\tt www-data} and has read permission for the group, or (ii) has world-readable UNIX permissions.

Using utilities for copying directories between systems can silently alter these DAC permissions in unintended ways, leading to serious security lapses. We illustrate this scenario using Apache {\tt httpd} and migration of its data using {\tt tar}. To study the impact of name collisions on the security parameters, we assume that the migration happens from a case-sensitive to a case-insensitive file system. The behavior of {\tt tar} discussed below draws from the discussion of \autoref{tab:tooling}.

To protect sensitive directories, {\tt httpd} can be configured to only allow authenticated users to access specific directories. A commonly used approach is to configure authentication via the {\tt .htaccess} file~\cite{httpd-auth} which lists the valid users/groups allowed to access a specific directory over {\tt HTTP}. All sub-directories inside the sensitive directory are also protected. We show that the use of additional security-oriented files can be exploited under the presence of name collisions.

\paragraph{Scenario}

\begin{figure}[ht]
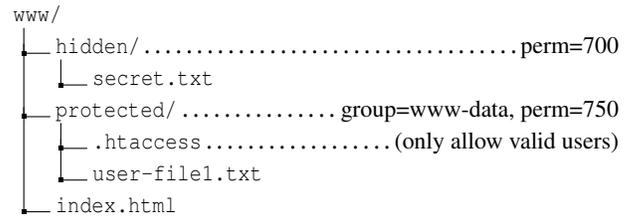

\small
\dirtree{%
 .1 www/.
 .2 hidden/\DTcomment{perm=700}.
 .3 secret.txt.
 .2 protected/\DTcomment{group=www-data, perm=750}.
 .3 .htaccess\DTcomment{(only allow valid users)}.
 .3 user-file1.txt.
 .2 index.html.
}
\vspace{-4pt}\caption{{\tt www/} on case-sensitive file system}
\vspace{-0.1in}
\label{fig:httpd-case}
\end{figure}

{\tt httpd} serves the contents of {\tt www/} (of~\autoref{fig:httpd-case}) over {\tt HTTP}. Initially, {\tt www/} is stored on a case-senstive file system. The directory {\tt hidden/} is inaccessible over {\tt HTTP} since the {\em others} permissions are cleared. Next, {\tt protected/} is configured to be accessible only to specific users using the {\tt .htaccess} file.

\paragraph{Adversary}
A UNIX user called Mallory has read-write access to {\tt www/} directory. However, DAC permissions prevent her from accessing {\tt hidden/} directory because its owner is another user. Additionally, {\tt protected} is inaccessible since Mallory does not belong to the group {\tt www-data}.

\begin{figure}[ht]
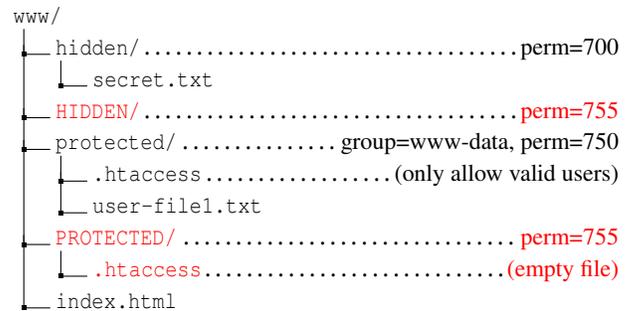
\vspace{-4pt}
\small
\newcommand{\adv}[1]{\textcolor{red}{#1}}
\dirtree{%
 .1 www/.
 .2 hidden/\DTcomment{perm=700}.
 .3 secret.txt.
 .2 \adv{HIDDEN/}\DTcomment{\adv{perm=755}}.
 .2 protected/\DTcomment{group=www-data, perm=750}.
 .3 .htaccess\DTcomment{(only allow valid users)}.
 .3 user-file1.txt.
 .2 \adv{PROTECTED/}\DTcomment{\adv{perm=755}}.
 .3 \adv{.htaccess}\DTcomment{\adv{(empty file)}}.
 .2 index.html.
}
\vspace{-4pt}\caption{Adversary modified {\tt www/} on the case-sensitive file system}
\vspace{-12pt}
\label{fig:httpd-case-adv}
\undef\adv
\end{figure}

She modifies {\tt www/} as shown in \autoref{fig:httpd-case-adv} and adds the {\tt HIDDEN/} and {\tt PROTECTED/} directories with the intent of gaining access to {\tt hidden/} and {\tt protected/} via a name collision.

\paragraph{Vulnerability}

\begin{figure}[ht]
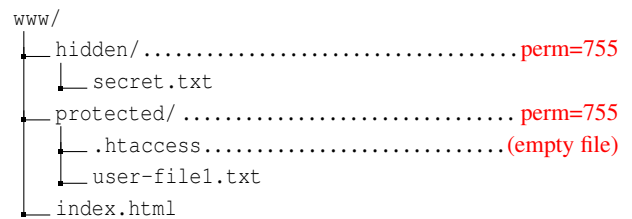
\vspace{-4pt}
\small
\newcommand{\adv}[1]{\textcolor{red}{#1}}
\dirtree{%
 .1 www/.
 .2 hidden/\DTcomment{\adv{perm=755}}.
 .3 secret.txt.
 .2 protected/\DTcomment{\adv{perm=755}}.
 .3 .htaccess\DTcomment{\adv{(empty file)}}.
 .3 user-file1.txt.
 .2 index.html.
}
\vspace{-4pt}\caption{{\tt www/} after migrating to case-insensitive file system}
\vspace{-10pt}
\label{fig:httpd-case-after-copy}
\undef\adv
\end{figure}

{\tt tar} is used to migrate the adversary-modified {\tt www/} directory to another system that uses a {\em case-insensitive} file system. \autoref{fig:httpd-case-after-copy} shows the state of the file system once the tarball (archive format of {\tt tar}) is extracted.

Now, the previously inaccessible {\tt hidden/} directory is now accessible over {\tt HTTP}. Additionally, since the {\tt .htaccess} file is cleared, unauthenticated users will be allowed to view {\tt protected/} over {\tt HTTP}.

\paragraph{Reporting}
We have reported this scenario to the Apache maintainers, but have not yet reached a resolution.
Using \autoref{tab:tooling}, we can reason about the above problems. Under a {\em directory -- directory} collision, {\tt tar} incorrectly modifies metadata. This happens for the {\tt hidden/} -- {\tt HIDDEN/} collision. Here, DAC permissions of the latter are applied to the former resulting in the leakage of secret files.

For {\em directory -- directory} collisions, {\tt tar} will also merge contents of both directories. For {\tt protected/} -- {\tt PROTECTED/} collision, this merger results in the empty {\tt .htaccess} file overwriting the original one that restricts access to authorized users. The end result is that all users are now allowed access to the new {\tt protected/} directory.

\section{Potential Defenses}
\label{sec:defense}

As discussed in the context of name confusion attacks in general in \autoref{subsec:relwork}, it can be difficult to produce defenses to prevent name collisions as well.  In this section, we discuss some options and their limitations.  


Name collisions are due to differences in case folding rules among file systems, e.g., case sensitivity and encodings, so it is difficult to ensure that name collisions cannot happen.  Suppose a system has only one file system.  Even then, an archive constructed on another file system using conflicting case folding rules may cause name collisions to occur when copying the archive.  Since user-space programs cannot determine the case-folding rules that may be applied to a file, user-space solutions alone will be unreliable.  In addition, they may be prone to TOCTTOU attacks~\cite{bishop-dilger,mcphee74}.  Thus, extending library calls like {\tt realpath} to detect name collisions will not sufficiently solve the problem.  In addition, system solutions lack knowledge of the programmer intent that caused the collision, so as for name confusion defenses, systems defenses will suffer from one or more of the follow limitations: it must have side information about the programs it protects, it must protect only a subset of all programs, it must be vulnerable to DoS attacks, it must have false-positives, or it must fail to prevent some  exploits~\cite{cai09oakland}. 

For example, one idea may be to write a wrapper to vet archives prior to expansion operations (e.g., {\tt tar} and {\tt zip}) to validate that each file in the archive will result in distinct file after expansion.  One way to do this is to check for name collisions due to case among all the files in the archive.  Although the notion that no files in an archive should collide with any other files within the archive seems intuitively reasonable, there are critical drawbacks to implementing such a defense.  For example, there may be files already in the target directory that may result in collisions, limit its utility.  More fundamentally, the case folding rules applied by such a wrapper are not guaranteed to be the same as those of the target directory.

As a result, we envision that defenses for name collisions will evolve in a manner similar to defenses for name confusions that utilize the {\tt open} commands.  Consider that the {\tt open} command has flags to check whether a file of a corresponding name exists a creation time, only opening that file when created anew (i.e., {\tt O\_CREAT|O\_EXCL}).  This call prevents a name collision from overwriting an existing file, but it may be too strong a defense.  Suppose one really wants to overwrite files of the same name, but prevent name collisions from modifying files that actually have differing names (i.e., that only match due to case folding).  In this case, a new flag is necessary, such as {\tt O\_EXCL\_NAME}, which prevents opening a file when the names differ (e.g., based on {\tt strcmp}), but not when such names match.

Unfortunately, even with variants of the {\tt open} command and other defenses, such as FileProvider classes in Android, programmers continue to make mistakes that lead to errors and vulnerabilities.  The challenge is for programmers to determine the intent of their operation, understand the threats faced in such an operation, and configure these complex, low-level commands in such a way that they block the threats while satisfying the intent.  Until file system APIs enable this combination of requirements, errors will remain common.

\section{Related Work}
\label{sec:relwork}

Researchers have proposed system and program defenses to thwart name confusion attacks for alias and squat cases.  To date, no defenses that are specific to name collisions have been proposed.

\paragraph{System Defenses}
Researchers have long known about name confusion attacks~\cite{mcphee74,bishop-dilger} and have proposed a variety of system defenses~\cite{chari2010you,cowan2001raceguard,dean2004fixing,openwall,park2004rps,payer2012protecting,lhee2005detection,tsafrir2008portably,tsyrklevich2003dynamic,uppuluri2005preventing,pu2006methodical}.  In a system defense, the operating system aims to enforce an invariant that prevents name confusion attacks from succeeding.  A challenge is that whether an operation is a feature or a vulnerability depends on the programmer's intent.  As a result, system defenses cannot prevent attacks completely without introducing the possibility of false positives~\cite{cai09oakland}.  Hybrid defenses have also been proposed~\cite{pf13eurosys,jigsaw14usenix} where the operating introspects into the process to leverage program state along with file system state in enforcement.  While reducing false positives, these techniques still lack programmer intent explicitly, resulting in some false positives.  

\paragraph{Program Defenses}
As a result, systems provide APIs for programmers to decide how to handle name confusion attacks.  Several file system APIs include flags to avoid using symbolic links entirely (e.g., {\tt O\_NOFOLLOW flag} for the {\tt open} system call), but in many cases programmers want to be able to use symbolic links.   Researchers have proposed program-specific defenses to configure APIs or program frameworks for preventing name confusion attacks~\cite{krohn2007information,porter2009operating,shapiro1999eros,watson2010capsicum}.  More advanced commands for file allow programmers to manage {\em how} files are open, including the impact of symbolic links.  For example, the {\tt openat} system call enables the user to open a directory first to validate its legitimacy before opening the remaining path.  {\tt openat2}
explicitly constrains how name resolution is performed to reduce the potential for attacks.

\section{Conclusion}
Interactions among file systems with differing encoding/case-sensitivity design decisions can lead to name collisions when performing maliciously crafted, or even ostensibly benign, copy operations. In this paper, we explored the impact that these name collisions can have on file system security. Current operating systems do not directly prevent name collision-based attacks, delegating that responsibility to programmers. In investigating the utilities used to copy file system resources and repositories/archives, we demonstrate that they often allow unsafe name collisions and lack the sort of uniformity in name-collision handling against which safer use policies could be easily crafted. Further, we show that many applications rely on potentially unsafe use of these utilities, opening themselves up to exploitable vulnerabilities. We examine three case studies in greater detail and demonstrate concrete vulnerabilities to name collisions. Finally, we suggest a series of directions for future research to systematically defend against name collision attacks.


\nocite{*}

\bibliographystyle{plain}
\bibliography{bibliography}


\clearpage
\pagebreak
\pagenumbering{roman} 
\setcounter{page}{1}
\appendix

\section{Command-line Flags for Utilities}
\label{sec:cmd-flags}

\subsection{tar}
\texttt{tar} can be used to copy files/directories across different systems. In the first step, \texttt{tar} reads files and directories hierarchies at a given source and creates an archive (also known as a tarball) that contains information about the source contents. Subsequently, \texttt{tar} can recreate the source at another target using the previously created archive. The archive is portable and can be moved across computer systems.

To generate \autoref{tab:tooling}, we used {\tt GNU tar (v1.30)} with the following command-line flags.
\begin{description}
    \item[-cf] create new archive and specify its name
    \item[-x] extract contents from the archive
    \item[-v] verbose mode (does not impact behavior)
\end{description}

\subsection{zip}
\texttt{zip} is similar to \texttt{tar}. It also creates an intermediate archive to copy files across computer systems. The following flags were used with \texttt{zip (v3.0)} to produce \autoref{tab:tooling}.

\begin{description}
    \item[-r] recursively traverse all directories (during archive creation)
    \item[--symlinks] store symbolic links in zip archive
\end{description}

\subsection{cp and cp*}
\texttt{cp} is a commonly used utility to copy files from a source to a target. Both the source and target need to be accessible on the same machine.

Unlike \texttt{tar} and \texttt{zip}, \texttt{cp} does not create an intermediate archive. The files and directories are directly created at the target by reading the source.

The following flags were used with {\tt cp (GNU Coreutils v8.30)} to generate \autoref{tab:tooling}.
\begin{description}
    \item[-a] do not deference symlinks, recursively copy directories, and preserve attributes (mode, ownership, timestamps, xattr, hardlinks and context)
\end{description}

{\tt cp} and {\tt cp*} use the same executable binary. The difference is in the way the command-line arguments are passed to the binary. Specifically, the format of specifying the source directories is different.

Consider that the source directory (to be copied) is {\tt foo}. For {\tt cp}, we will pass it as {\tt foo/} while for {\tt cp*} we will use {\tt foo}. Note the trailing {\tt /} is missing in the latter case. Just this difference changes the behavior of {\tt cp} as noted in \autoref{tab:tooling}.

\subsection{rsync}
\texttt{rsync} can be used to copy, transfer, or sync file/directories on the same machine, or across machines (over a custom \url{rsync://} protocol).

The following flags were used with {\tt rsync (v3.1.3)} to generate \autoref{tab:tooling}.
\begin{description}
    \item[-a] recursively copy directories, preserve symlinks, timestamps, DAC permissions, owners, groups, device files and special files
    \item[-H] preserve hardlinks
\end{description}

\end{document}